\documentclass[12pt]{article}
\usepackage{amssymb}
\usepackage{epsfig}

\newcommand{\bq}{\begin{equation}}
\newcommand{\eq}{\end{equation}}
\newcommand{\bqa}{\begin{eqnarray}}
\newcommand{\eqa}{\end{eqnarray}}
\newcommand{\ben}{\begin{enumerate}}
\newcommand{\een}{\end{enumerate}}
\newcommand{\bc}{\begin{center}}
\newcommand{\ec}{\end{center}}

\def\bqb{\begin{eqnarray*}}
\def\eqb{\end{eqnarray*}}
\def\gsim{\gtrsim}
\def\lsim{\lesssim}
 
\def\pr#1#2#3{ Phys. Rev. ${\bf{#1}}$ (#2) #3}

\def\pl#1#2#3{ Phys. Lett. ${\bf{#1}}$ (#2) #3}
\def\prep#1#2#3{ Phys. Rep. ${\bf{#1}}$ (#2) #3}
\def\np#1#2#3{ Nucl. Phys. ${\bf{#1}}$ (#2) #3}
\def\zp#1#2#3{ Z. f. Phys. ${\bf{#1}}$ (#2) #3}

\def\ie{{\it i.e.\/}}
\def\eg{{\it e.g.\/}}

\global\nulldelimiterspace = 0pt

\def\vb#1{{\bf #1}}

\def\L{ {\cal L }}
\def\O{ {\cal O }}

\def\T{ {\cal T }}
\def\R{ {\cal R }}

\def\swd{s^2_W}
\def\cwd{c^2_W}

\def\mw{M_W}
\def\mz{M_Z}

\def\mh{m_H}

\def\vtau{\mbox{\boldmath $\tau$}}

\begin{document}
\thispagestyle{empty}

\begin{raggedleft}
THES-TP 97/03\\
March 1997\\
hep-ph/9703446\\
\end{raggedleft} 
\vspace*{3cm}
\begin{center}
{\Large\bf{ Signatures of CP-violation in
$\gamma \gamma \rightarrow H$\\
\vspace{0.25em}
using polarized beams}}\footnote{Partially supported by the EC
contract CHRX-CT94-0579.\\ 
Email: georgia@ccf.auth.gr}\\
\vspace{2em}
{\large G.J. Gounaris and G.P. Tsirigoti}
\vspace{1em}\\
Department of Theoretical Physics, University of
Thessaloniki, \\GR 54006, \hspace{0.25em} Thessaloniki, Greece\\
\vspace*{2cm}
{\bf Abstract}\hspace{2.2cm}\null
\end{center}

The possibility of observing CP-violation in the process $\gamma
\gamma \rightarrow H$ is investigated for masses of the Higgs
particle in the interval 
$\mz \lsim  m_H \lsim 2 m_t$, using a 0.5TeV 
tunable linear $e^{+} e^{-}$ collider through laser backscattering. 
The use of polarized beams allows
the formation of two different asymmetries sensitive 
to CP-violating New Physics (NP) interactions among the gauge and 
Higgs bosons. It is shown that very low values
of the corresponding NP couplings can be probed, for a large range
of the Higgs mass.\\
\vspace*{0.5cm}


\setcounter{footnote}{0} 
\clearpage
\newpage

\section{Introduction}

High energy $e^{-}e^{+}$ linear colliders are crucial in
thoroughly investigating the Higgs sector of the Standard Model
(SM) and beyond. Their significance is twofold:

Assuming that the Higgs particle really exists and its mass is
below 1TeV, it 
is generally believed that Higgs production at LHC or in $e^{-}e^{+}$
collisions, through $W^+W^-$ fusion or $e^{-}e^{+}\to HZ$, will give
a good signal leading  to the discovery of
the Higgs boson \cite{Workshop, Zerwas}. 

Once the Higgs boson is detected and its mass is measured, 
one would like to test whether its properties are as
predicted by SM. This means measuring its width and its 
interactions with  the matter and gauge fields. In this respect,
it is necessary to check whether New Physics (NP) 
beyond SM exists, which induces new Higgs 
interactions. The $e^-e^+$ Linear Colliders, applied either
directly or in their
$\gamma \gamma$ mode\footnote{We assume here that such a 
$\gamma \gamma$ Collider will some day be feasible.} , provide a very
useful machinery for such studies. Since in SM the $\gamma\gamma H$
coupling arises only at the one loop level and it is mediated by
loops of all charged particles with non zero mass, measurement of
$\gamma \gamma \to H$ can reveal the existence of possibly new
interactions induced by new heavy particles
that cannot be directly produced in these next generation colliders
\cite{Gunion}.  

In the present work we assume that no new particles
responsible for the New Physics (NP) beyond SM, will be
producible in the
future colliders. Moreover, we assume that the scale 
$\Lambda_{NP}$ of NP is
sufficiently large and that the Higgs particle really exists. 
Under these conditions, NP may be described in 
terms of $dimension=6$  $SU(3)\times SU(2)
\times U(1)$ gauge invariant
operators creating new CP conserving and CP violating
couplings \cite{Buchmuller}. 
A complete list of such operators inducing purely
bosonic CP conserving couplings among  the Higgs and the gauge
bosons  can be looked at \cite{Hag1, DeR, papad}. \par

On the other hand, the complete list of the $dim=6$ purely
bosonic CP-violating and $SU(3)\times SU(2)
\times U(1)$ gauge invariant operators, 
 may be represented as 
\bqa
\tilde{\O}_W &= & {1\over3!}~  \epsilon_{ijk}~ W^{i\mu\nu}
  W^{j}_{\nu\lambda} \tilde{W}^{k\lambda}_{\ \ \ \mu} \ \ \
 ,  \ \  \label{Wtil} \\
\tilde{\O}_G &= & {1\over3!}~  f_{ijk}~ G^{i\mu\nu}
  G^{j}_{\nu\lambda} \tilde{G}^{k\lambda}_{\ \ \ \mu} \ \ \
 ,  \ \ \   \label{Gtil} \\
\tilde{\O}_{WW} & = &  (\Phi^\dagger \Phi )\,    
\vb{W}^{\mu\nu} \cdot \tilde{\vb{W}}_{\mu\nu} \ \
\  ,  \ \ \label{WWtil}\\
\tilde{\O}_{BB} & = &  (\Phi^\dagger \Phi ) B^{\mu\nu} \
\tilde{B}_{\mu\nu} \ \ \  , \ \ \   \label{BBtil} \\[0.1cm] 
\tilde{\O}_{GG} & = &  (\Phi^\dagger \Phi)\, 
\delta_{ij} G^{i \mu\nu} \cdot  \tilde{G}^j_{\mu\nu} \ \
\  .  \ \    \label{GGtil} \\  
\tilde{\O}_{BW} & =& \frac{1}{2}~ \Phi^\dagger B_{\mu \nu}
{\vtau} \cdot \tilde{\vb{W}}^{\mu \nu} \Phi
\ \ \  , \ \  \label{BWtil}  
\eqa
where \eg\@ ${\tilde \vb{W}_{\mu\nu}}=\frac{1}{2}
\varepsilon_{\kappa\lambda\mu\nu} \vb{W}^{\kappa\lambda}$.
This list of operators differs
from the one in \cite{Stong, Choi} in the respect that 
we have included 
the gluonic  operators $\tilde{\O}_G$ and 
$\tilde{\O}_{GG}$  and omitted  
\bqa
\tilde{\O}_{W\Phi} & = & i\, (D_\mu \Phi)^\dagger 
{\vtau} \cdot \tilde{\vb{W}}^{\mu \nu} (D_\nu \Phi) \ \ \  , \ \
\label{WPhitil} \\
\tilde{\O}_{B\Phi} & = & i\, (D_\mu \Phi)^\dagger 
\tilde{B}^{\mu \nu} (D_\nu
\Phi)\ \ \  , \ \ \label{BPhitil}
\eqa
since they are related to the operators in
(\ref{Wtil}-\ref{BWtil}) through
\bqa
\tilde{\O}_{W\Phi} & = & \frac{g}{4}\ \tilde{\O}_{WW}+
\frac{g\prime}{2}\ \tilde{\O}_{BW} \ \ , \\
\tilde{\O}_{B\Phi} & = & \frac{g\prime}{4}\ \tilde{\O}_{BB}+
\frac{g}{2}\ \tilde{\O}_{BW} \ \ ,
\eqa
imposed by the Bianchi identities $D^\mu {\tilde \vb{W}_{\mu\nu}}=0$.
In (\ref{WPhitil}, \ref{BPhitil}), $D_\nu$ is the usual gauge
covariant derivative. Various dynamical scenarios for the arising of
the operators (\ref{Wtil}-\ref{BWtil}), have been discussed in
\cite{dyn, papad}. Motivated from these we note in particular, 
that the gluonic operators 
$\tilde{\O}_G$ and $\tilde{\O}_{GG}$ are easily 
generated whenever the heavy particles inducing NP are
coloured.\par 

Concerning the list (\ref{Wtil}-\ref{BWtil}),
we remark that  $\tilde{\O}_W$
and $\tilde{\O}_{BW}$ are the only operators involving triple
electroweak gauge boson couplings (TGC), while  $\tilde{\O}_{WW}$,
$\tilde{\O}_{BB}$ and $\tilde{\O}_{GG}$  contain only
Higgs NP interactions and no TGC. Instead of this
parameterization, another in principle equivalent
one could had been given by the list in  
(\ref{Wtil}, \ref{Gtil}, \ref{GGtil}-\ref{BPhitil}).
We prefer the first parameterization though, because it clearly
separates  the triple gauge
boson couplings (TGC) possibly induced by NP, from the Higgs 
involving interactions, which could also be generated. 
Because of their ability to induce TGC, the operators
$\tilde{\O}_W$ and $\tilde{\O}_{BW}$ are already quite strongly
constrained by the existing information on the neutron and
electron electric dipole moments \cite{DeR-CP}. More precisely,
only one combination of these two operators is in fact
constrained by the
electric dipole moments, but  these constraints 
may be further improved (for both operators) 
in the future, by $e^-e^+ \to WW$ studies through 
polarized beams in the Next Linear Collider (NLC) at 500GeV and
above \cite{Papadopoulos}. \par

Our aim here is to concentrate on possible NP effects involving
only  the Higgs interactions with the electroweak gauge bosons.
We are of course aware of the fact that naturality, 
together with the use of the list of operators in 
(\ref{Wtil}, \ref{Gtil}, 
\ref{GGtil}-\ref{BPhitil}) as an NP basis, have been
invoked to argue that the present very strong
constraints on the CP violating TGC, would suggest that it would
be very improbable to have any non-vanishing Higgs-gauge boson NP
couplings also \cite{DeR, DeR-CP}. Nevertheless, we  
feel that a direct measurement of such interactions is still  
very useful and important. Particularly because  
in the case studied here, where it is assumed that no new
particles would be 
producible in the future; the study of the underlying nature and
interactions of the obscure Higgs particle 
seems to be a prime candidate for supplying crucial hints on NP.\par    

In   $e^-e^+$, 
$\gamma \gamma $ and $e\gamma$ collisions producing
gauge and/or Higgs bosons, both, the CP-conserving  
and the CP-violating  boson interactions contribute  
\cite{Zerwas, Higgs-R, Boudjema1}. In order to be able to
disentangle the CP violating Higgs interactions, 
suitable processes and polarization effects must be looked at 
\cite{Choi, Hag2, Kramer, Vlachos}. 
Such CP violating purely electroweak NP couplings are
described above by the operators
$\tilde{\O}_{WW}$ and $\tilde{\O}_{BB}$. In principle, it should
be possible to disentangle these two operators  by studying suitable
polarization effects and angular dependencies in
$e^-e^+ \to H\gamma,~HZ$ \cite{Stong, Kramer, Vlachos},
as well as in $\gamma \gamma \to WW $ \cite{Belanger, Choi},
and in single Higgs production.\par

In this paper we study $\gamma \gamma \to H$ for suitably polarized
laser and $e^\pm$ beams and construct two asymmetries
which are sensitive to CP violating Higgs gauge boson interactions. 
These asymmetries are analogous to those used in \cite{Choi} 
for studying $\gamma \gamma \to WW$. It turns out that they are 
both sensitive to the same combination of the $\tilde{\O}_{WW}$ and
$\tilde{\O}_{BB}$ couplings. The high photon luminosities
expected in these machines though, combined with a reasonably 
expected adequate understanding of the background,
guarantees that a very high sensitivity to this coupling 
combination, should be possible. Augmenting  
this information with the one obtained from 
$e^-e^+ \to H\gamma,~HZ$, will allow a thorough study of all 
possible CP violating Higgs involving 
interactions, induced
at the level of $dim=6$ gauge invariant operators.\par

\section{Polarization asymmetries sensitive to CP violation in  
$\gamma\gamma \to H$.}

Polarization effects in the process $\gamma \gamma \to H$
provide a very efficient way of disentangling the operators
$\tilde{\O}_{WW}$ and $\tilde{\O}_{BB}$ from the rest of the
CP-conserving and the CP-violating $dim=6$
$SU(3)\times SU(2)\times U(1)$ gauge invariant interactions;
(compare (\ref{WWtil}, \ref{BBtil})). 
The effective Lagrangian describing the part of NP induced by
these operators is 
\bq
\label{LNPCP}    
\L_{NP}~ = ~ \frac{\bar d}{v^2}\,
\tilde{\O}_{WW}+\frac{{\bar d}_{B}}{v^2}\, \tilde{\O}_{BB}
\ \ , 
\eq
where $v=2\mw/g \simeq 246GeV$. From this we calculate the
NP contribution to the $\T_{\mu_1 \mu_2}$ amplitude 
for $\gamma \gamma \to H$, where $\mu_1$ and $\mu_2$ are
the helicities of the two incoming photons\footnote{The phase of 
$\T$ is defined by its connection to the ${\cal S}$-matrix 
as ${\cal S}=1+i\T$.}. We remark that CP-transformation 
implies for this amplitude that
\bq
\label{CP-helicity}
\T_{\mu_1  \mu_2}~=~\pm \T_{-\mu_2 ~ - \mu_1} \ \ ,
\eq
where the upper (lower) sign is valid for  the  
CP-conserving (CP-violating) part of the amplitude.\par  

In Appendix A we give the density matrix  $\R$ 
(in the helicity basis),  of the two colliding
photons, produced by backscattering of two laser beams from the 
incoming highly energetic $e^\pm$. Using (\ref{R-matrix}), we write 
\bq
\label{sigggH}
\sigma(\gamma \gamma \to H) =  \left 
\{\frac{d\L^{\gamma \gamma}(\tau)}{d\tau}\right \}_{\tau=\tau_H}
 \frac{\pi}{s_{ee} \mh^2 }
\sum_{\mu_j} \T_{\mu_1 \mu_2}\T_{\mu_1\prime \mu_2\prime}^*
\ \langle \rho^{BN}_{\mu_1\mu_1\prime} 
\bar{\rho}^{BN}_{\mu_2 \mu_2\prime} \rangle \ \ ,
\eq
where the $\gamma \gamma$ luminosity has been defined
through (\ref{Lgammagamma}, \ref{laser-C}, \ref{laser-D}), 
$\tau_H\equiv \mh^2/s_{ee}$,
while the normalized density matrices of the two backscattered
photons are given by (\ref{rhoBN}) as
\bqa
\rho^{BN} & =& \frac{1}{2}
\left (\matrix{1+\xi_2(x) & - \xi_{13}(x) e^{-2i\varphi} \cr
- \xi_{13}(x)  e^{+2i\varphi} & 1-\xi_2(x) } \right ) \ \ ,
\nonumber \\[.3cm]
\bar \rho^{BN} & =& \frac{1}{2}
\left (\matrix{1+\bar \xi_2(x) & - 
\bar \xi_{13}(x) e^{2i\bar \varphi} \cr
- \bar \xi_{13}(x) e^{-2i\bar \varphi} & 1-\bar \xi_2(x) }
\right )  \ \ .
\eqa 
In the definition of the azimuthal angle $\bar \varphi$
for the second photon in the last equation, 
the sign has been changed to take care of
the fact that we choose to define it not around its own
momentum, but rather around the momentum of the 
oppositely moving photon.\par

The $\gamma \gamma \to H$ cross section defined in (\ref{sigggH}) is then
given as 
\bqa
\label{Sigma}
&& \sigma (\gamma \gamma \to H) ~ = ~
\left 
\{\frac{d\L^{\gamma \gamma}(\tau)}{d\tau}\right \}_{\tau=\tau_H}
\cdot \Bigg [(1+\langle\xi_2 \bar \xi_2 \rangle) \Sigma_{unp}
\nonumber \\
&&+  \langle \xi_{13} \bar \xi_{13} \rangle \cos[2(\varphi -\bar
\varphi)] \Sigma_1
+\langle \xi_{13} \bar \xi_{13} \rangle \sin[2(\varphi -\bar
\varphi)] \Sigma_2 +
\langle \xi_2 +\bar \xi_2 \rangle \Sigma_3 \Bigg ] \ ,
\eqa
where the averages of (the products of) the $\xi_j$ parameters,
determining the density matrices of the backscattered photons, 
have been defined
through (\ref{xi-averages1}, \ref{xi-averages2},
\ref{R-matrix}). In (\ref{Sigma}) we have  used the
definitions  
\bqa
\Sigma_{unp} & = & \frac{\pi}{4 s_{ee} \mh^2}~\{ |\T_{++}|^2
+ |\T_{--}|^2 \} \ \ , \label{S-unp} \\
\Sigma_1 & =& \frac{\pi}{2 s_{ee} \mh^2}~ Re (\T_{++}\T_{--}^*)
=\Sigma_{unp} \ \ , \label{S1} \\
\Sigma_2 & =& \frac{\pi}{2 s_{ee} \mh^2}~ Im (\T_{++}\T_{--}^*)
\ \ , \label{S2} \\  
\Sigma_3 & = & \frac{\pi}{4 s_{ee} \mh^2}~\{ |\T_{++}|^2
- |\T_{--}|^2 \} \ \ . \label{S3} 
\eqa \par

According to (\ref{CP-helicity}), only $\Sigma_2$ and $\Sigma_3$
are sensitive to the CP violating NP couplings to linear order.
Thus, to 1-loop order in the SM contribution, and to linear order
in the CP violating NP couplings of (\ref{LNPCP}), we have
\bqa
\Sigma_{unp}& = &
\frac{G_F \mh^2 \alpha^2}{16\sqrt{2}\pi s_{ee}}~
\Bigg |\frac{4}{3}F_t+ F_W \Bigg |^2 \ \ , \label{S-unpSM} \\
\Sigma_1 & = &
\frac{G_F \mh^2 \alpha^2}{16\sqrt{2}\pi s_{ee}}~
\Bigg |\frac{4}{3}F_t+ F_W \Bigg |^2 \ \ , \label{S1-SM}\\
\Sigma_2 & = & \frac{- G_F \mh^2 \alpha}{\sqrt{2} s_{ee}}~
Re \left (\frac{4}{3}F_t +F_W \right )(\bar d \swd +\bar d_B \cwd)
\ \ , \label{S2-NP} \\
\Sigma_3 & = & \frac{- G_F \mh^2 \alpha}{\sqrt{2} s_{ee}}~
Im \left (\frac{4}{3}F_t +F_W \right )(\bar d \swd +\bar d_B \cwd)
\ \ , \label{S3-NP}
\eqa
where $F_t$ and $F_W$, which denote the top and $W$ loop SM
contributions,  can be found in 
\cite{HgammaSM, Gunion, Higgs-R}. \par

The expected annual number of events for double laser 
backscattering in $e^-e^+$ 
colliders, is obtained by multiplying the cross-section 
in (\ref{sigggH}) with the annual luminosity
$\L_{ee}\simeq 20fb^{-1} \mbox{year}^{-1}$  
for a $0.5TeV$ collider. Therefore  
\bq
\label{number}
N_{\tau_H} = {\cal L}_{ee}\ \sigma(\gamma \gamma \to H)
\ \ .
\eq
We next construct the two possible
CP-odd asymmetries. The first may be observable whenever 
there are  non-vanishing 
average linear polarization parameters $\xi_{13}$ and $\bar \xi_{13}$
for the two colliding  photons. It is obtained by performing
measurements for two different values of the angle
$\chi\equiv \varphi - \bar\varphi$ between the linear
polarizations of these photons. It is given by 
\bq
\label{Alintilde}
\tilde A_{lin} ~ = ~\frac{|
N_{\tau_H}(\chi=\frac{\pi}{4})-N_{\tau_H}(\chi=-
\frac{\pi}{4})|}{N_{\tau_H}(\chi=\frac{\pi}{4})
+N_{\tau_H}(\chi=- \frac{\pi}{4})} ~ = ~
\frac{\langle \xi_{13}\bar \xi_{13}\rangle }
{1+\langle \xi_{2}\bar \xi_{2}\rangle }~ A_{lin} \ \ , \\
\eq
with
\bq
\label{AlinNP}
A_{lin}=\frac{|\Sigma_{2}|}{\Sigma_{unp}}\ \ .
\eq
As seen from (\ref{AlinNP}), $A_{lin}$ is determined through 
(\ref{S-unpSM}, \ref{S2-NP}) solely by the CP-violating NP
induced quantity $\Sigma_2$.   \par

The construction of the second CP violating asymmetry is
possible whenever there is a non vanishing average for the sum
of the circular polarizations $\langle \xi_2 + \bar \xi_2 \rangle $
of the two back scattered photons. This requires large average
helicities for both, the laser and the $e^\pm$ beams, 
(see the Appendix). Using (\ref{Sigma}) and  remarking
that the sign of $\langle \xi_2+\bar \xi_2 \rangle $ changes whenever 
the signs of $(P_e, ~P_\gamma)$ and $(\bar P_e, ~\bar P_\gamma)$
are simultaneously changed, we construct the asymmetry
$\tilde A_{circ}$ by making measurements with
two opposite values of these polarization pairs. We
thus get 
\bq
\label{Acirctilde}
\tilde A_{circ} ~=~ \frac{|N_{\tau_H}^{++}-N_{\tau_H}^{--}|}
{N_{\tau_H}^{++}+ N_{\tau_{H}}^{--}}~ =~
\frac{\vert \langle \xi_2+\bar \xi_2\rangle \vert}
{1+\langle \xi_{2}\bar 
\xi_{2}\rangle }~ A_{circ} \ \ ,
\eq
with 
\bq
\label{AcircNP}
A_{circ} ~=~ \frac{|\Sigma_3|}{\Sigma_{unp}} \ \ .  
\eq\par

Each one of the asymmetries $\tilde A_{lin}$ and 
$\tilde A_{circ}$ is a product of two factors:
The first one originates from the degree of polarization of the
two photons building the collider, while the second one depends 
on the product of an SM 1-loop contribution and 
another contribution sensitive to the CP violating NP  
interactions. This second factor is denoted respectively by
$A_{lin}$ and $A_{circ}$, and satisfies
$A_{lin}(SM)=A_{circ}(SM)=0$ in the SM case. To study 
observability limits for the NP couplings from the use 
of these asymmetries, we need the expressions for the expected
1-standard deviation statistical uncertainties for each of them in the
SM case. These  are
\bqa
\delta A_{lin}(SM) & =  & \frac{|1+\langle \xi_2 \bar \xi_2 \rangle|}
{\sqrt{\left (N_{\tau_H}(\chi=\frac{\pi}{4})
+N_{\tau_H}(\chi=- \frac{\pi}{4}\right )}\langle
\xi_{13} \bar \xi_{13} \rangle } \  \ \ , \label{dAlin} \\
\delta A_{circ}(SM) & = &\left \vert \frac{1+\langle \xi_2 \bar \xi_2 \rangle}
{\sqrt{\left (N_{\tau_H}^{++} +N_{\tau_H}^{--}\right )} \langle
\xi_2 + \bar \xi_2 \rangle } \right \vert \  \ \ . \label{dAcirc}
\eqa

We also note from (\ref{S2-NP}, \ref{S3-NP}), that both
asymmetries measure the same combination $\bar d \swd +\bar d_B
\cwd$ of the NP couplings called $\bar{d}_{\gamma \gamma}$ in
\cite{Vlachos}. \par

\section{Testing CP violation  in $\gamma\gamma\rightarrow
H$ through an  $e^{-}e^{+}$ collider.}

As explained in the Appendix, the polarized photons needed to
study the CP violating
contributions to $\gamma \gamma\rightarrow H$, 
are obtained by double laser backscattering from the $e^{-},~e^{+}$
beams \cite{laser, Boudjema1}. The outgoing photons are
produced almost in the same direction with $e^{-},~e^{+}$. 
All relevant 
formulae for the description of the Compton scattering kinematics 
in this framework, are  collected in the Appendix.
Here we only note that these are  
characterized by two dimensionless
parameters, $x_0=4E\omega_0/m_e^2$ and
$x=\omega/E$, where E is the  energy of
the electron beam, $\omega_0$ is the laser energy and $x$ is
the fraction of the $e^{-}$ beam energy carried away by the
final photon. The maximum value of $x$ is determined by
$x_0$, through the relation $x_{max}=x_0/(1+x_0)$.
Operation of the collider below the $e^{-}e^{+}$ pair
production threshold sets an upper limit to the value of
$x_0$, so that $x_0\leq 2(1+\sqrt{2})$. Thus the final
photons can take, as much as  $\sim 83\%$ of the electron
beam energy; compare (\ref{laser-kin}). 
A lower limit in the value of $x_0$ is also set for
each specific process under consideration, by the
masses of the particles produced in the process 
studied \cite{Hag2}. Therefore,  the
allowed range for $x_0$, for a given $e^-e^+$
center of mass energy is 
\bq
\label{x0-bounds}
{\sum\limits_{i=1}^n m_{i}\over \sqrt{s}-\sum\limits_{i=1}^n
m_{i}}\leq x_0\leq 2(1+\sqrt{2}) \ \ , 
\eq
where the sum includes all produced masses in a 
$\gamma\gamma$ collision. \par

There are various  general options for  operating the
photon-photon collider \cite{laser, Boudjema1}:
\begin{itemize}
\item 
The conversion point (C.P.) where the Compton
backscattering occurs, may be a  few cm away from
the interaction point (I.P.) where the $\gamma\gamma$ collisions
take place. Since the most energetic photons are those with the
smallest scattering angle, it is only those that finally manage
to reach the I.P. Therefore, the further away from the
I.P. the conversion occurs, the more monochromatic 
the photon beam becomes, (but with some loss in the integrated
luminosity of course). This
particular set up can be very advantageous for processes
where production of some resonance occurs, whose mass is a
priori known. In this case, the collider may be tuned, 
so as to operate
in a narrow window around the relevant specific value of the invariant
$\gamma\gamma$ mass \cite{Richard, Boudjema2}. This way, 
there is also the
possibility of reducing the background in cases where the
cross-section of the main process dominates. 
\item 
Choosing a configuration where I.P. and C.P. coincide, 
makes the photon spectrum
rather flat in the unpolarized and the $P_eP_\gamma >0$ cases, 
which may be useful  for searching particles
with unknown masses. It may help also in the simultaneous study of
more than one processes, dominating at different regions of
the invariant $\gamma\gamma$ energies.  
\end{itemize}

We turn now to the specific properties of the process  
$\gamma\gamma \to H$ in the Next Linear Collider. Assuming 
that the mass of the Higgs boson has been
measured before, in the $e^{+}e^{-}$ mode or perhaps at LHC, it 
may  be possible to tune the parameters of the collider 
so that it operates, (to some extent), in a narrow window around
the Higgs mass.  
The use of polarized beams in searching for any CP violating new
interactions among the gauge and Higgs
bosons is investigated in this section 
\cite{Gunion, Kramer, Barger, Boudjema2}. The measurements of the
two quantities $A_{lin}$ and $A_{circ}$, which mostly require 
different polarization conditions, are considered separately.

\bigskip

A. {\bf Measurement of $A_{lin}$ }

As seen from (\ref{Alintilde}), the measurement of $A_{lin}$
requires laser beams with some linear polarization $P_t$.
Then, the resulting photons building the
photon-photon collider are also polarized in the same
direction. The degree of linear polarization transferred 
to them, $\xi_{13}$, depends on the parameter 
$x_0$ (determining the maximum collider energy) and on $P_t$.

This Stokes parameter $\xi_{13}$ 
is given in (\ref{laser-xi13}), 
as a function of its fractional energy $x$ 
\cite{laser}. As explained in the Appendix and 
shown in Fig.3b, the degree of linear polarization $\xi_{13}$ 
is very small for low energy photons, while it tends to its maximum
value $\xi_{13max}=2(1+x_0)/[1+(1+x_0)^2] P_{t}$, when
$x\rightarrow x_{max}$. It is obvious from this, that
in order to increase the linear polarization transferred from the 
laser beams to
high energy photons, it is best to have a machine design such
that $x_0$ is as small as possible. However, when $x_0$
decreases, the highest Higgs mass producible through 
$\gamma \gamma \to H$ in a given collider, also decreases. 
The best choice should therefore be decided by tuning 
the collider after the Higgs particle
discovery and the measurement of its mass.\par         

Before discussing the actual measurement of $A_{lin}$, we should
comment a bit about the background for detection of $\gamma
\gamma \to H \to X $.
In the case that $\mh \lsim 150GeV$, in which
$H \to b \bar b $ dominates, the main background process is
$\gamma\gamma \to b \bar b$. This background however, is strongly
peaked in the forward-backward direction, while the Higgs decay is 
isotropic in the Higgs frame. It may also
be interesting to  utilize
laser beams which are partly circularly and partly linearly
polarized, like \eg\@ $P_\gamma =P_t=1/\sqrt 2$; 
compare (\ref{laser-rho}, \ref{PtPgineq}).
Because then the backscattered photons acquire circular as well
as linear polarizations, which may be useful in
further reducing the $\gamma \gamma \to b \bar b$ background; 
since its dependence of  on the Stokes
parameters is very different from the Higgs mediated 
contribution given in (\ref{Sigma}). More specifically, the
background is strongly suppressed for $\langle \xi_2 \bar \xi_2
\rangle = +1$. Concerning the magnitude of this background,
a detailed discussion can be found in 
\cite{gunion-haber} where there is also a
plot of the minimum and maximum masses for which the number of
predicted Higgs signal events (S) and background events (B) are
such that $S \gsim 10$ and $ S/\sqrt {B}\gsim 5$.
We note that these conditions are satisfied for  
Higgs masses in the  range $100-150 GeV$, 
for all possible degrees of polarization \cite{gunion-haber}.
 
On the other hand, for the Higgs mass range $200 \lsim \mh\lsim 350GeV$
for which the relevant Higgs decay mode is $H \to
ZZ \to l^-l^+X$, we note that the background should be very
small, since  there is no SM tree level contribution to
the $\gamma\gamma \to ZZ$ continuum. Combining this with
a $ZZ$ detection efficiency of $\sim 18\%$, it is concluded
in \cite{Richard} that the Higgs study in 
the above mass range is determined by the absolute event 
rate only. \par

Using therefore (partially) linear polarizations we should be able to 
measure the $\tilde A_{lin}$ asymmetry defined in 
(\ref{Alintilde}), which in turn determines
$A_{lin}$ and the combination $\bar d \swd +\bar d_B \cwd$ 
of the NP couplings; (compare  (\ref{AlinNP})).
To get a feeling on the possible limits that can be established,
we plot in Fig.1 the ratio
\bq
\label{NSDAlin}
\mbox{NSD for } A_{lin} ~\equiv ~ \frac{A_{lin}}{\delta A_{lin}(SM)}
\ \ ,
\eq
where (\ref{AlinNP}, \ref{dAlin}) should be used. This ratio
describes number times the measurable 
factor $A_{lin}$, exceeds its expected
statistical fluctuation in SM, for various choices of the NP couplings
\cite{Belanger, Hag2, Choi}. Fig.1 presents $\mbox{NSD for } A_{lin}$ for 
a $\sqrt{s_{ee}}=0.5TeV$ collider, using the small value of 
$x_0=0.5$, as motivated above.  \par

The sensitivity to the NP coupling combination 
$\bar d \swd +\bar d_B \cwd$, which can be reached by studying 
$A_{lin}$, depends on the Higgs mass. It can be obtained from 
Tables I and
II, where the NP couplings inducing a $3\sigma$ effect are
tabulated, using   an integrated annual luminosity 
$\bar {\cal L}_{ee} \simeq 20 fb^{-1}\mbox{year}^{-1}$
for a 0.5TeV  Collider and various values 
for $x_0$ \cite{Richard}.
As can be seen from these tables, the sensitivity to the above NP
couplings generally increases as $x_0$ decreases. 

The results in Table I apply for $100 \lsim \mh \lsim 150GeV$
and were derived on the basis of the 
$H \to b \bar b$ mode for which a $25\%$ detection efficiency
is assumed, \cite{Richard, Workshop, Zerwas}. 
In fact, this is also what determines the overall 
number of the  $(\varphi -\bar\varphi)$ averaged events in the
last column of the Table I. Thus, if $100 \lsim \mh \lsim 150GeV$,
then limits on the NP coupling 
$\bar{d}_{\gamma \gamma} \equiv \bar d \swd +\bar d_B \cwd$
are at the level $10^{-3}-10^{-4}$ seem possible. In Table I
we give results for $P_t=1$ as well as for $P_t=1/\sqrt 2$
(in parentheses), in order to give a feeling of the possible
implications from using laser 
beams with a partial circular polarization\footnote{Remember the
preceding discussion concerning the 
$b \bar b$ background.}. \par

On the other hand, the results in Table II apply for 
$200 \lsim \mh \lsim 350GeV$ and were derived on the basis of
the $H \to ZZ \to l^-l^+X$. In both Tables I and II, it is
always checked that 
the number of expected events for the
chosen decay channels is of the order of a hundred,  for 
all Higgs masses considered and a $0.5TeV$ collider.

For the high $\mh$ part in Table II, we should also remark 
that as the Higgs mass increases, the
sensitivity to the NP CP-violating couplings is reduced. The
reason is that the lower limit for the $x_0$ parameter
also increases and consequently the degree of linear polarization
transfer decreases; (compare (\ref{x0-bounds})). 
On top of this,  there is a further
reduction of sensitivity, since $A_{lin}$, being proportional to the
real part of SM contribution, decreases as $\mh$ increases.
However, as seen from Table II for Higgs masses in the region 
 200-250GeV, sensitivity limits on the NP couplings like 
$(\bar d \swd +\bar d_B \cwd) \sim 10^{-3}-10^{-2}$ can be
obtained  at the $2\sigma$ or $3\sigma$ level, from 
measurements of $A_{lin}$.
 
\bigskip

B. {\bf Measurement of $A_{circ}$}        

Eqs. (\ref{Acirctilde}, \ref{laser-xi2}) indicate that the 
measurement of $A_{circ}$ require the existence of circularly
polarized photons, which may be obtained by backscattering 
similarly polarized laser beams from polarized $e^\pm$.
Both, the energy spectrum of the resulting photons and the degree of the 
polarization transferred, depend on the way we choose the initial
average helicities, as well as the conversion and interaction points; 
(see Appendix and \cite{laser, Belanger, Boudjema1}).
If C.P. and I.P. coincide, this spectrum is rather flat
for $P_e P_\gamma > 0$, and peaked towards the higher energies for 
$P_e P_\gamma < 0$; (compare Fig.3c). 
 
As for  the $A_{lin}$ case, the statistical significance 
of a possibly non vanishing value induced by NP to  $A_{circ}$, 
is again given the number of times this asymmetry exceeds 
the expected statistical uncertainty in SM; \ie
\bq
\label{NSDAcirc}
\mbox{NSD for } A_{circ} ~=~ \frac{A_{circ}}{\delta A_{circ}(SM)}
\ \ ,
\eq
where (\ref{AcircNP}, \ref{dAcirc}) should be used. 
Note that $A_{circ}(SM)=0$.\par

The numerical study of (\ref{Acirctilde}) indicates that 
the measurement of 
$A_{circ}$ depends very sensitively on the average 
polarizations along their momenta of the $e^\pm$ beams
$P_e,~\bar P_e$, and of the laser photons 
$P_\gamma , ~ \bar P_\gamma $. On the other hand, 
since the Higgs particle has no spin, in order to enhance 
its production we must choose the same average 
helicities in both arms of the collider; \ie\@ 
$P_e= \bar P_e$ and $P_\gamma =\bar P_\gamma $.   
If a configuration is selected in which the conversion and
interaction points coincide, then we could choose 
$P_e P_\gamma=+1$, where  the
backscattered  photons acquire for most of the energy range,
a mean helicity of the same sign as the initial ones, and only
very near to the maximum energy this helicity changes 
sign; (see Fig.3c in the Appendix). In
this way the best polarization transfer is achieved (nearly
$100\%$) for almost the entire range of the invariant
$\gamma\gamma$ masses, where the luminosity is also 
significant.\par

If, however, a distance is put between C.P. and I.P., then
choosing the polarizations such that 
$P_e P_\gamma= \bar P_e \bar P_\gamma =-1$ enhances the
production of photons with the highest energies, leading thus to a gain
in luminosity. Assuming that the Higgs mass is known and   
tuning   the collider so that the most
energetic photons are the ones contributing to Higgs production,
allows a most efficient use of their high circular polarization which 
facilitates the measurement of $A_{circ}$. This is what is done 
in Fig.2a below.\par    

It is also important observe from (\ref{AcircNP}, \ref{S3-NP}),
that $A_{circ}$ is proportional to the imaginary part of the 
SM contribution to the $\gamma \gamma \to H$ amplitude,
which is very small for $m_H \lsim 2\mw $. 
This imaginary part starts becoming appreciable only above the
$WW$-threshold. Therefore, a measurement of $A_{circ}$ can be
useful  only for
$\mh \gsim 2\mz$, where the most useful  Higgs decay mode is 
$ H \to ZZ \to l^{+}l^{-}X$ decay.
Thus in Fig.2, we plot $\mbox{NSD for } A_{circ}$
for $2M_{z} \lsim m_{H} \lsim 350 GeV$ and various values of the  
NP coupling combination $\bar d \swd +\bar d_B \cwd$.\par

Fig.2a corresponds to $P_eP_\gamma=-1$, while
Fig.2b corresponds to $P_eP_\gamma=+1$. The other parameters in
these figures are chosen so that they facilitate the measurement
of the $A_{circ}$ asymmetry.  Thus
for the case of Fig.2a, where the spectrum of the backscattered
photons is peaked towards the high energy side, we use
the highest value of $x_0$ possible, namely $x_0=4.82$, as well
as collider whose energy is tuned to the Higgs mass
like  $\sqrt{s_{ee}}=\mh/0.75$.
On the contrary for Fig.2b, corresponding to a rather flat
spectrum of backscattered photons, we do not  tune $s_{ee}$
to the Higgs mass, but we just fix  $\sqrt{s_{ee}}=0.5TeV$
and $x_0=4$ \cite{Richard}.
The plots in Fig.2ab assume that  C.P. and I.P. coincide.\par

In Tables III and IV we also give the $3\sigma$ sensitivity
limits to the above couplings from $A_{circ}$, as a function of
$\mh$, using the above polarization choices. In these tables
we fix the  energy to $\sqrt s_{ee}=0.5 TeV$, and vary
$x_0$. It can be concluded
from these Tables that sensitivity limits like 
 $(\bar d \swd +\bar d_B \cwd)\sim 10^{-4}$, should be possible
 in the whole range $2M_{z} \lsim m_{H} \lsim 350 GeV$. \par

\section{Final discussion}
 
We have shown in the present work that if the Higgs particle
predicted by SM really exists, then the study of the process 
$\gamma \gamma \to H $ using polarized beams in Next Linear
Colliders, provide a very sensitive test for the  
existence of any CP-violating interactions among the gauge and
 Higgs bosons. Various polarization configurations for the 
$e^\pm$ beams and the laser photons give complementary
information and provide consistency checks for the study of such 
couplings. In particular, if the Higgs mass is in the
range of (100-150)GeV, then linearly polarized laser beams 
may be used in a $\sim 0.5TeV~e^{-}e^{+}$ collider, in order to
measure the asymmetry $\tilde A_{lin}$ defined
above, which is sensitive to CP-violating NP induced interactions. 
This way sensitivity limits on the CP violating NP coupling
$\bar d \swd +\bar d_B \cwd$ may be obtained at 
the level of $10^{-3}-10^{-4}$; 
(compare Table I). \par

If the Higgs particle is in the  range 
$2\mz \lsim m_{H} \sim 200GeV$, then the most efficient way to
look for CP violating NP interactions, is through
measurement of the $\tilde A_{circ}$ asymmetry, using 
circularly  polarized beams. In this case, non vanishing 
average helicities for the electron
as well as the laser beams 
are  necessary, in order  to have good sensitivity to the 
anomalous couplings. We have also seen that in case 
$P_eP_\gamma>0$, it may be useful to try a tunable
Linear Collider. Thus, also for these higher masses, 
the sensitivity limits on the NP coupling $\bar d \swd +\bar d_B
\cwd$ are again at the $10^{-4}$ level; (see Table III, IV). \par

In both cases, it should be possible to
disentangle the CP violating forces, from the CP conserving ones
affecting the same processes.\par

From the theoretical point of view, a nonzero value for any
of these anomalous couplings provides, through unitarity, a hint
on the related NP scale $\Lambda_{NP}$. Therefore, it is interesting to
translate the above sensitivity limits for the couplings, to 
corresponding lower bounds on NP scales. Assuming  that
only one of the operators $\tilde{\O}_{WW}$ or 
$\tilde{\O}_{BB}$ acts at a time,  the unitarity requirement
gives the relations \cite{Vlachos}
\bqa
|\bar d| & \simeq  & \frac{104.5
(M_W/\Lambda_{NP})^2}{1+3(M_W/\Lambda_{NP})}
\ \  \ , \label{uni-dbar} \\
|\bar d_{B}| & \simeq &
\frac{195.8(M_W/\Lambda_{NP})^2}{1+100(M_W/\Lambda_{NP})^2} \ \ .
\label{uni-dBbar}
\eqa   
Using then a non vanishing value for the corresponding NP 
coupling at the level of the aforementioned sensitivity
limits, we  calculate from the unitarity relations  
the scale $\Lambda_{NP}$ where unitarity is first reached. 
In general $\Lambda_{NP}$ depends on the operator
considered. Its value provides a rough estimate of the energy
scale where either new strong
interactions will develop,  or  new particles 
will be produced.   From (\ref{uni-dbar}) and the above 
sensitivity limits, we would   conclude that  
the study of  $\gamma \gamma \to H$ at a $0.5TeV$
collider  can probe NP scales in the range of 10-20TeV,
in case NP is generated by $\tilde{\O}_{WW}$. 
Because of (\ref{uni-dBbar}), this scale increases to 
30-50TeV, if we assume that the NP
forces are due to the operator $\tilde{\O}_{BB}$.\par

In summary, using polarized beams for realizing 
the $\gamma \gamma$ colliders, it is possible to
construct observables sensitive only to the 
~CP-violating NP couplings, and thereby distinguish
them  from the CP conserving
ones. This is not attainable if  unpolarized beams are
only used. The overall conclusion
for the $dim=6$ gauge invariant NP interactions 
considered above, is that single Higgs production in $\gamma \gamma$
collisions at $0.5TeV$ tunable linear Collider, can be used to put
 limits on the NP coupling  
$\bar d \swd +\bar d_B \cwd$ at the $10^{-3}-10^{-4}$ level, 
for Higgs masses in the ranges
$100 \lsim \mh \lsim 150 GeV$ and $200 \lsim \mh \lsim 350 GeV$.
This information is complementary, and at least an order of
magnitude  more precise than the one attainable through
production   of $WW$ pairs in $\gamma \gamma $ 
collisions, where the same combination of NP couplings
is measured \cite{Choi, Higgs-H, Higgs-R}. 
Information on independent combinations of the CP violating
couplings at the level of $10^{-2}$
may be obtained by looking at $e^-e^+ \to H\gamma,~HZ$
\cite{Vlachos, Stong, Kramer}. Thus, a combination of such
measurements should be able to constrain separately each of the two
CP violating couplings $\bar d$ and $\bar d_{B}$ at
the $10^{-2}$ level.

\newpage
\renewcommand{\theequation}{A.\arabic{equation}}
\setcounter{equation}{0}
\setcounter{section}{0}

{\large \bf Appendix A : Density matrix of backscattered 
photon.}

Following \cite{laser}, we collect in this appendix 
the formulae describing the helicity density
matrix of the photon produced by backscattering a laser photon
from an incoming highly energetic $e^\pm$ beam.\par
 
We denote by $E$ the energy of the incoming $e^\pm$ beam, 
while $P_e=2\lambda_e$
describes its polarization along its momentum, and $\lambda_e$ 
is its average helicity. The $e^\pm$ beam is assumed to 
collide with a laser 
photon moving along the opposite direction with energy 
$\omega_0$ and characterized, in the basis of  
its helicity states,  by the normalized density 
matrix \cite{Berestetskii}
\bq
\label{laser-rho}
\rho_{laser}^N ~ = ~\frac{1}{2}
\left (\matrix{1+P_\gamma & -P_t e^{-2i\varphi} \cr
-P_t e^{+2i\varphi} & 1-P_\gamma } \right ) \ \ . 
\eq
Here  $P_\gamma$ describes the average helicity of the laser photon
and $P_t$ denotes its  maximum average linear polarization 
occurring of course along a direction perpendicular
to the photon  momentum, described by the azimuthal angle
$\varphi$ (with respect to this momentum). By definition 
\bq
\label{PtPgineq}
0 \leq P_\gamma^2+P_t^2 \leq 1 \ \ .
\eq\par

After the Compton scattering of $e^\pm$ from the laser photon, 
the electron beam looses most of its energy and a beam of "backscattered
photons" is
produced, moving essentially along the direction of the original
$e^\pm$ momentum and characterized, in its helicity basis, 
by the density matrix 
\bqa
\rho^B &=& \frac{dN}{dx}~\rho^{BN} \ \ , \label{rhoB} \\
\rho^{BN} & =& \frac{1}{2}
\left (\matrix{1+\xi_2(x) & - \xi_{13}(x) e^{-2i\varphi} \cr
- \xi_{13}(x)  e^{+2i\varphi} & 1-\xi_2(x) } \right ) \ \ ,
\label{rhoBN} 
\eqa 
where $x \equiv \omega/E$ and $x_0 \equiv 4E \omega_0/m_e^2$; with
$\omega$ being the energy of the back-scattered photon,
and $\omega_0$, $E$ have been defined above. 
These satisfy the kinematical constraints
\bq
\label{laser-kin}
0 \leq x \leq x_{max} ~\equiv~ \frac{x_0}{1+x_0} \ \ \ \ , 
\ \ \ \ 0\leq x_0 \leq 2 (1+\sqrt 2) \ \ ,
\eq
which implies  that the backscattered photon can take up to 83\%
of the $e^\pm$ energy.
In analogy to (\ref{PtPgineq}), the elements of $\rho^{BN}$
also  satisfy 
\bq
0 \leq \xi_2^2(x)+ \xi_{13}^2(x) \leq 1 \ \ .
\eq\par

In (\ref{rhoB}), $dN/dx$ is the number of
backscattered photons per unit of $x$, normalized to a unit of 
flux for the 
incoming $e^\pm$ beam; while $\rho^{BN}$ is the normalized 
photon density matrix, ($Tr\rho^{BN}=1$). 
We note from (\ref{rhoBN}, \ref{laser-rho}),  that the 
azimuthal angles of the
maximum average linear polarizations of the backscattered 
and the laser photons, defined around their respective momenta, 
are the same. The functions appearing 
in (\ref{rhoB}, \ref{rhoBN}), which determine the spectrum of
the backscattered photon immediately after its production at the
{\it conversion point},
are given by \cite{laser} 
\bqa
\frac{dN(x)}{dx} &= & \frac{C(x)}{D(x_0)}\ \ , \label{gamma-flux} \\
C(x) &=& f_0(x)+P_eP_\gamma f_1(x) \ \ , \label{laser-C} \\
D(x_0) & = & D_0 (x_0) + P_eP_\gamma D_1 (x_0) \ \ ,
\label{laser-D} \\
\xi_2(x) & = & \frac{P_e f_2(x) +P_\gamma f_3(x)}{C(x)} \ \ , 
\label{laser-xi2}\\
\xi_{13}(x) & =& \frac{2 r^2 P_t}{C(x)} \ \ , \label{laser-xi13} 
\eqa
where
\bqa
f_0(x) & = &  \frac{1}{1-x}+1-x-4r(1-r) \ , \ \\
f_1(x) &= & \frac{x}{1-x}~(1- 2r)(2-x)\ , \\
f_2(x) &= & x_0 r [1+(1-x)(1-2r)^2]\ , \\
f_3(x) &=& (1-2r)\left (\frac{1}{1-x}+1-x \right ) \ , \\  
r(x) & = & \frac{x}{x_0 (1-x)} \ \ ,
\eqa
and
\bqa
D_0 (x_0)=\int_0^{x_{max}} dx f_0(x) & =& [1-  \frac{4}{x_0}
- \frac{8}{x_0^2}]\ln(1+x_0) +\frac{1}{2}+\frac{8}{x_0}-
 \frac{1}{2(1+x_0)^2}  , \label{laser-D0} \\
D_1 (x_0)~=~\int_0^{x_{max}} dx f_1(x) & =
& [1 + \frac{2}{x_0} ]\ln(1+x_0) -\frac{5}{2}+\frac{1}{1+x_0}-
 \frac{1}{2(1+x_0)^2}  , \label{laser-D1}
\eqa
where $x_{max}$ is defined in (\ref{laser-kin}). \par

The elements of density matrix of the backscattered photon, for
various choices of the $e^\pm$ polarization $P_e$, and the laser
parameters $x_0$ and $(P_\gamma,~ P_t)$, are presented in
Fig.3abc. Thus, in Fig.3a, we give the backscattered photon flux
$dN/dx$ as a function of $x$. As seen from
(\ref{gamma-flux}-\ref{laser-D}), $dN/dx$ depends only on the 
product $P_eP_\gamma$ and the parameter $x_0$ determining the
highest value of $x$ through (\ref{laser-kin}). In Fig.3b, the
average linear polarization $\xi_{13}$ of the backscattered
photon is shown. As seen from comparing
Fig.3ab, the demand of a high linear polarization would favour
a small value of $x_0$, which has the drawback that 
the highest energy of the
back scattered photon decreases. Finally, in Fig.3c, we present 
the average
circular polarization $\xi_2$ of the back scattered photon as a
function of $x$, for various choices of $P_e$, $P_\gamma$ and
$x_0$. As seen in Fig.3c, the $x$-dependence of 
$\xi_2$ is quite sensitive to the relative sign of $P_e$,
$P_\gamma$. \par

If an $e^-e^+$ Collider is transformed to $\gamma \gamma$ one by
using two identical lasers, then the (unnormalized) density matrix
$\R_{\mu_1\mu_2 ; \mu_1\prime \mu_2\prime}$ of the $\gamma
\gamma$-pair in their helicity basis, is related to  the
$\rho^B$, $\bar{\rho}^B$ matrices by (compare (\ref{rhoB})) 
\bqa
\label{R-matrix}
\frac{d}{d\tau}\, 
\R_{\mu_1\mu_2 ; \mu_1\prime \mu_2\prime}(\tau)
& = & \rho^B_{\mu_1\mu_1\prime} \bigotimes 
\bar{\rho}^B_{\mu_2 \mu_2\prime} \equiv
\int_{\frac{\tau}{x_{max}}}^{x_{max}}
\frac{dx}{x} \rho^B_{\mu_1\mu_1\prime}(x) \bar{\rho}^B_{\mu_2
\mu_2\prime} \left (\frac{\tau}{x}
\right )\ , \nonumber \\
&\equiv & \frac{d\L^{\gamma \gamma}(\tau)}{d\tau}\ \langle
\rho^{BN}_{\mu_1\mu_1\prime} 
\bar{\rho}^{BN}_{\mu_2 \mu_2\prime} \rangle \ \ ,
\eqa
where 
\bq
\tau ~ \equiv ~ \frac{s_{\gamma \gamma}}{s_{ee}} \ \ ,
\eq
with $s_{ee}$ and $s_{\gamma \gamma}$ being the squares of the
c.m. energies of the $e^-e^+$ and $\gamma \gamma $ systems
respectively. In the r.h.s. of (\ref{R-matrix}),
$d\L^{\gamma \gamma}/d\tau$ is the overall $\gamma
\gamma $ luminosity per unit $e^-e^+$ flux,  defined
by the convolution of the separate $\gamma $ luminosities
appearing in (\ref{gamma-flux}-\ref{laser-D}).   
If the {\it conversion} points
where each of the two photons are produced through laser backscattering, 
coincide with their 
{\it interaction} point, then 
\bq
\label{Lgammagamma} 
\frac{d\L^{\gamma \gamma}}{d\tau} ~=~\frac{1}{D^2(x_0)}
\int_{\frac{\tau}{x_{max}}}^{x_{max}}
\frac{dx}{x} C(x) C\left( \frac{\tau}{x} \right ) \ .
\eq
We note from (\ref{laser-C}, \ref{laser-D}), that the overall
flux of the backscattered photons depends on $P_e$ as well as on
the circular polarization of the laser photons used to produce them.
The definition of $d\L^{\gamma \gamma}/d\tau$ 
 specifies also the definition of the "averages" $\langle
\rho^{BN}_{\mu_1\mu_1\prime} \bar{\rho}^{BN}_{\mu_2 \mu_2\prime}
\rangle$ appearing in the r.h.s. of 
(\ref{R-matrix}) for the two photons.
These averages  may also be used to define 
$\langle \xi_2\bar \xi_2 \rangle) $,
$\langle \xi_{13} \bar \xi_{13} \rangle $,
$\langle \xi_2 + \bar \xi_2 \rangle $, using the form of 
(\ref{rhoBN}). Thus,
\bqa
\langle \xi_i \bar \xi_j \rangle & = &
\frac{(C \xi_i \bigotimes C \xi_j )}{C \bigotimes C} \ \ , 
\label{xi-averages1}\\
\langle \xi_2 + \bar \xi_2 \rangle & = &
\frac{(C (\xi_2 +\bar \xi_2) \bigotimes C )}{C \bigotimes C} \ \
\label{xi-averages2}.
\eqa \par

The above assumption that the {\it conversion}
and {\it interaction} points coincide, may most   probably
not be imposed when the $\gamma \gamma $ collision experiments will
be designed. Even in such a case though,
the main  modification we would need to make in the preceding
formulae is to appropriately increase the lower limit in the
integrals in (\ref{R-matrix}, \ref{Lgammagamma}). Thus,
increasing the distance between the {\it conversion} and {\it
interaction} points, will tend to select only the highest energy
part of the spectrum for each of the beams of the two 
colliding photons. \par


\renewcommand{\arraystretch}{1.2}
\begin{table}
\begin{small}
\begin{tabular}{|c|c|c|c|} \hline
\multicolumn{4}{|l|}{{\bf TABLE I.} $3\sigma$ upper bounds on
CP-violating NP couplings from $A_{lin}$ asymmetry, using}\\ 
\multicolumn{4}{|c|}{$H \to b \bar b$ for the two polarization
choices:}\\
\multicolumn{4}{|c|}{$P_t=\bar P_t=1, ~P_\gamma=\bar P_\gamma =0$, 
 \hspace*{0.5cm} ($P_t=\bar P_t=P_\gamma=\bar P_\gamma= 1/\sqrt 2, 
~P_e=\bar P_e=1$).}\\
\hline\hline
\multicolumn{1}{|c|}{$m_{H}$ (GeV) } &
\multicolumn{1}{|c|}{$x_0$} & \multicolumn{1}{|c|}{upper limit
on ${\bar d_{B}} c_{W}^{2}+\bar d s_{W}^{2}$} &
\multicolumn{1}{|c|}{$(\varphi-\bar \varphi)$-averaged Events }\\ \hline
 & 0.5 & $10^{-3}$  (2x$10^{-3}$) & 197 (168) \\ \cline{2-4}
100 & 0.8 & 2x$10^{-2}$  (1.6x$10^{-2}$) & 184 (233) \\ \cline{2-4}
 & 1 & 0.12  (0.05) & 160 (255) \\ \hline
 & 0.5 & 5x$10^{-4}$  (8.5x$10^{-4}$) & 200 (160) \\ \cline{2-4}
 & 0.8 & 8x$10^{-3}$  (6x$10^{-3}$) & 207 (223) \\ \cline{2-4}
120 & 1 & 0.05  (0.017) & 192 (264) \\ \cline{2-4}
 & 1.5 & 0.16  (0.11) & 175 (306) \\ \hline
 & 0.5 & 4x$10^{-4}$  (6.5x$10^{-4}$) & 144 (111) \\ \cline{2-4}
 & 0.8 & 5x$10^{-3}$  (3.2x$10^{-3}$) & 159 (143) \\ \cline{2-4}
140 & 1 & 0.03  (0.01) & 156 (181) \\ \cline{2-4}
 & 1.5 & 0.09  (0.06) & 147 (237) \\ \cline{2-4}
 & 2 & 0.22  (0.22) & 138 (252) \\ \hline
 & 0.5 & 5x$10^{-4}$  (8x$10^{-4}$) & 85 (64) \\ \cline{2-4}
 & 0.8 & 4.3x$10^{-3}$  (3x$10^{-3}$) & 106 (87) \\ \cline{2-4}
150 & 1 & 0.023  (0.008) & 107 (113) \\ \cline{2-4}
 & 1.5 & 0.08  (0.05) & 102 (157) \\ \cline{2-4}
 & 2 & 0.2  (0.2) & 97 (174) \\ \hline 
\end{tabular}
\end{small}
\end{table}

\begin{table}
\begin{small}
\begin{tabular}{|c|c|c|c|} \hline
\multicolumn{4}{|l|}{{\bf TABLE II.} $3\sigma$ upper bounds on
CP-violating NP couplings from $A_{lin}$ asymmetry, using}\\
\multicolumn{4}{|c|}{ $H\to ZZ \to l^-l^+X$  with
$P_t=\bar P_t=1,~P_\gamma=\bar P_\gamma= 0$.}\\
\hline\hline
\multicolumn{1}{|c|}{$m_{H}$ (GeV) } &
\multicolumn{1}{|c|}{$x_0$} & \multicolumn{1}{|c|}{upper limit
on ${\bar d_{B}} c_{W}^{2}+\bar d s_{W}^{2}$ } &
\multicolumn{1}{|c|}{ $(\varphi-\bar \varphi)$-averaged  Events }\\ \hline
 & 0.8 & $10^{-3}$ & 76 \\ \cline{2-4}
 & 1 & 1.5x$10^{-3}$ & 88 \\ \cline{2-4}
200 & 1.5 & 7x$10^{-3}$ & 97 \\ \cline{2-4}
 & 2 & 0.022 & 102 \\ \cline{2-4}
 & 2.5 & 0.06 & 102 \\ \hline
 & 1 & 1.3x$10^{-3}$ & 51 \\ \cline{2-4}
230 & 1.5 & 3.7x$10^{-3}$ & 79 \\ \cline{2-4}
 & 2 & 1.1x$10^{-2}$ & 86 \\ \cline{2-4}
 & 2.5 & 0.027 & 89 \\ \hline
 & 1.5 & 2.8x$10^{-3}$ & 66 \\ \cline{2-4}
250 & 2 & 7.7x$10^{-3}$ & 72\\ \cline{2-4}
 & 2.5 & 0.018 & 77 \\ \hline
 & 1.5 & 2.7x$10^{-3}$ & 41 \\ \cline{2-4}
280 & 2 & 5.2x$10^{-3}$ & 55 \\ \cline{2-4}
 & 2.5 & 0.011 & 60 \\ \cline{2-4}
 & 4 & 0.074 & 67 \\ \hline
 & 2 & 4.8x$10^{-3}$ & 44 \\ \cline{2-4}
300 & 2.5 & 9.3x$10^{-3}$ & 51 \\ \cline{2-4}
 & 4 & 0.055 & 59 \\ \hline
 & 2 & 1.2x$10^{-2}$ & 84 \\ \cline{2-4}
330 & 2.5 & 9.8x$10^{-3}$ & 36 \\ \cline{2-4}
 & 4 & 0.044 & 48 \\ \cline{2-4}
 & 4.82 & 0.09 & 51 \\ \hline
 & 2.5 & 0.024 & 14 \\ \cline{2-4}
350 & 4 & 0.06 & 41 \\ \cline{2-4}
 & 4.82 & 0.11 & 45 \\ \hline
\end{tabular}
\end{small}
\end{table}

\begin{table}
\begin{small}
\begin{tabular}{|c|c|c|c|} \hline
\multicolumn{4}{|l|}{{\bf TABLE III.} $3\sigma$ upper bounds on
CP-violating NP couplings from $A_{circ}$ asymmetry, using}\\
\multicolumn{4}{|c|}{$H\to ZZ \to l^-l^+X$ decay 
and circularly polarized laser beams with
$P_e=\bar P_e =-P_\gamma=-\bar P_\gamma=\pm 1 $.} \\
\hline\hline
\multicolumn{1}{|c|}{$m_{H}$ (GeV) } &
\multicolumn{1}{|c|}{$x_0$} & \multicolumn{1}{|c|}{upper limit
on ${\bar d_{B}} c_{W}^{2}+\bar d s_{W}^{2}$} &
\multicolumn{1}{|c|}{$\xi_2$-averaged Events}\\ \hline
 & 1 & 5.7x$10^{-4}$ & 141 \\ \cline{2-4}
 & 1.5 & 6x$10^{-4}$ & 133 \\ \cline{2-4}
200 & 2 & 6.2x$10^{-4}$ & 121 \\ \cline{2-4}
 & 2.5 & 6.5x$10^{-4}$ & 112 \\ \cline{2-4}
 & 4 & 7x$10^{-4}$ & 95 \\ \cline{2-4}
 & 4.82 & 7.3x$10^{-4}$ & 89 \\ \hline
 & 1 & 6.3x$10^{-4}$ & 49 \\ \cline{2-4}
 & 1.5 & 4.2x$10^{-4}$ & 109 \\ \cline{2-4}
240 & 2 & 4.3x$10^{-4}$ & 105 \\ \cline{2-4}
 & 2.5 & 4.4x$10^{-4}$ & 99 \\ \cline{2-4}
 & 4 & 4.7x$10^{-4}$ & 85 \\ \cline{2-4}
 & 4.82 & 5x$10^{-4}$ & 80 \\ \hline
 & 1.5 & 4.5x$10^{-4}$ & 55 \\ \cline{2-4}
 & 2 & 3.7x$10^{-4}$ & 78 \\ \cline{2-4}
280 & 2.5 & 3.7x$10^{-4}$ & 78 \\ \cline{2-4}
 & 4 & 4x$10^{-4}$ & 70 \\ \cline{2-4}
 & 4.82 & 4x$10^{-4}$ & 66 \\ \hline
 & 2 & 4.8x$10^{-4}$ & 31 \\ \cline{2-4}
320 & 2.5 & 3.6x$10^{-4}$ & 53 \\ \cline{2-4}
 & 4 & 3.5x$10^{-4}$ & 56 \\ \cline{2-4}
 & 4.82 & 3.6x$10^{-4}$ & 54 \\ \hline
 & 2.5 & 6x$10^{-4}$ & 15 \\ \cline{2-4}
350 & 4 & 3.4x$10^{-4}$ & 45 \\ \cline{2-4}
 & 4.82 & 3.4x$10^{-4}$ & 46 \\ \hline
\end{tabular}
\end{small}
\end{table}

\begin{table}
\begin{small}
\begin{tabular}{|c|c|c|c|} \hline
\multicolumn{4}{|l|}{{\bf TABLE IV.} $3\sigma$ upper bounds on
CP-violating NP couplings from $A_{circ}$ asymmetry, using}\\ 
\multicolumn{4}{|c|}{ $H\to ZZ \to l^-l^+X$ decay 
and circularly polarized laser beams with
$P_e =\bar P_e = P_\gamma =\bar P_\gamma=\pm 1$.} \\
\hline\hline
\multicolumn{1}{|c|}{$m_{H}$ (GeV) } &
\multicolumn{1}{|c|}{$x_0$} & \multicolumn{1}{|c|}{upper limit
on ${\bar d_{B}} c_{W}^{2}+\bar d s_{W}^{2}$} &
\multicolumn{1}{|c|}{$\xi_2$-averaged Events}\\ \hline
 & 1 & 7x$10^{-4}$ & 94 \\ \cline{2-4}
 & 1.5 & 6x$10^{-4}$ & 128 \\ \cline{2-4}
200 & 2 & 6x$10^{-4}$ & 131 \\ \cline{2-4}
 & 2.5 & 6.1x$10^{-4}$ & 125 \\ \cline{2-4}
 & 4 & 6.8x$10^{-4}$ & 102 \\ \cline{2-4}
 & 4.82 & 7.1x$10^{-4}$ & 92 \\ \hline
 & 1 & $10^{-3}$ & 18 \\ \cline{2-4}
 & 1.5 & 5.3x$10^{-4}$ & 68 \\ \cline{2-4}
240 & 2 & 4.7x$10^{-4}$ & 86 \\ \cline{2-4}
 & 2.5 & 4.6x$10^{-4}$ & 92 \\ \cline{2-4}
 & 4 & 4.7x$10^{-4}$ & 85 \\ \cline{2-4}
 & 4.82 & 5x$10^{-4}$ & 80 \\ \hline
 & 1.5 & 8.1x$10^{-4}$ & 17 \\ \cline{2-4}
 & 2 & 5.3x$10^{-4}$ & 39 \\ \cline{2-4}
280 & 2.5 & 4.6x$10^{-4}$ & 51 \\ \cline{2-4}
 & 4 & 4.3x$10^{-4}$ & 59 \\ \cline{2-4}
 & 4.82 & 4.3x$10^{-4}$ & 58 \\ \hline
 & 2 & $10^{-3}$ & 6 \\ \cline{2-4}
320 & 2.5 & 6.3x$10^{-4}$ & 17 \\ \cline{2-4}
 & 4 & 4.5x$10^{-4}$ & 34 \\ \cline{2-4}
 & 4.82 & 4.4x$10^{-4}$ & 37 \\ \hline
 & 2.5 & 1.6x$10^{-3}$ & 2 \\ \cline{2-4}
350 & 4 & 5.5x$10^{-4}$ & 18 \\ \cline{2-4}
 & 4.82 & 5x$10^{-4}$ & 22 \\ \hline
\end{tabular}
\end{small}
\end{table}



\newpage

\begin{figure}[p]
\begin{center}
\mbox{
\epsfig{file=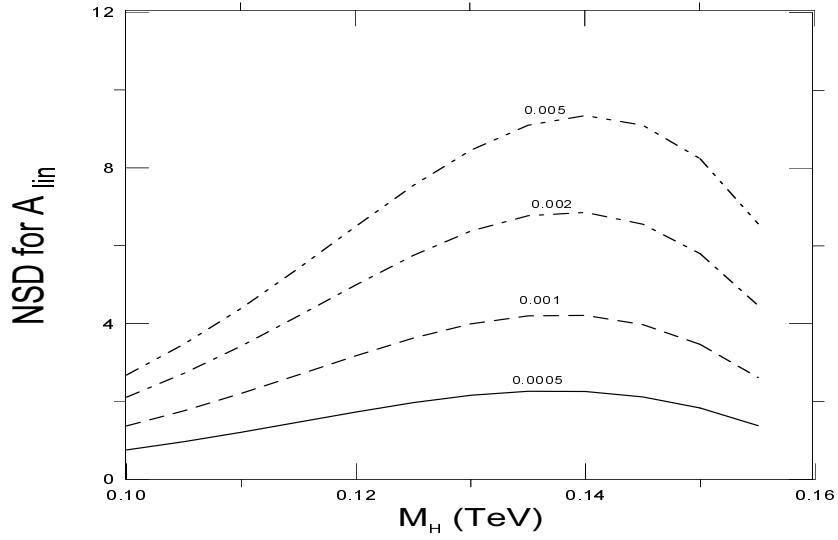,bbllx=1.5cm,bblly=0cm,bburx=18cm,bbury=27cm,
height=10cm,width=9cm}}
\vspace*{-2cm}
\caption[1]{ NSD for the 
$A_{lin}$-asymmetry, for various choices of the
($\bar d_B \cwd +\bar d \swd $) combination of couplings,
using $P_t=1$ and $x_0=0.5$ in a 
$\sqrt s_{ee}=0.5TeV$ Collider.}
\end{center}
\end{figure}

\clearpage
\newpage

\begin{figure}[p]
\begin{center}
\mbox{
\epsfig{file=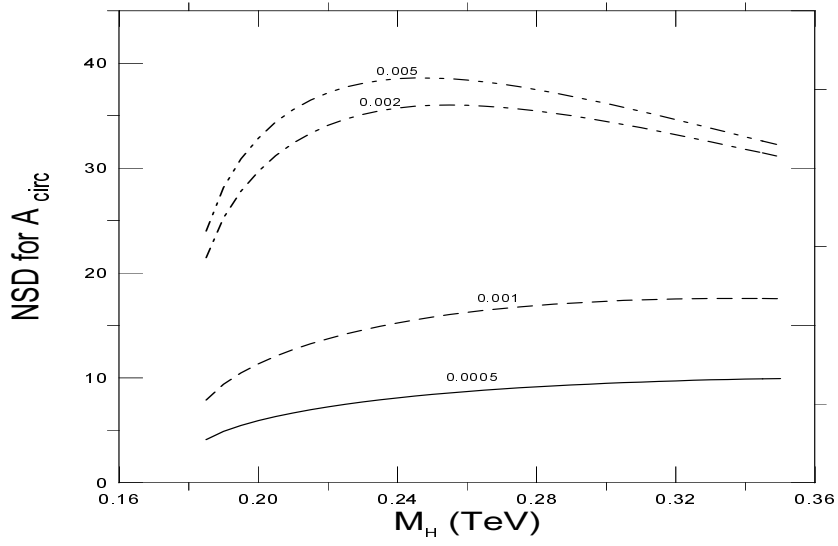,bbllx=1.5cm,bblly=0cm,bburx=18cm,bbury=27cm,
height=10cm,width=9cm}}
\vspace*{-2cm}
\end{center}
\hspace{8cm} (a)\\
\begin{center}
\mbox{
\epsfig{file=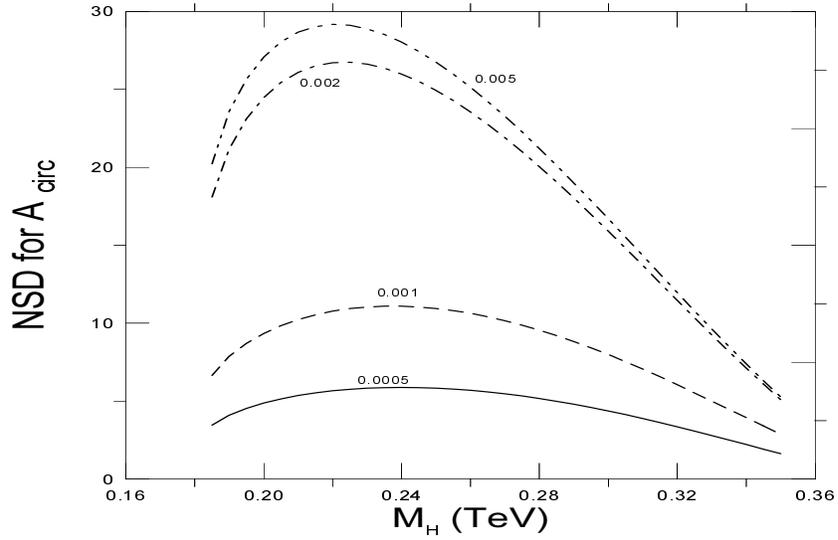,bbllx=1.5cm,bblly=0cm,bburx=18cm,bbury=27cm,
height=10cm,width=9cm}}
\vspace*{-2cm}
\end{center}
\hspace{8cm} (b)\\
\caption[2]{ NSD for the $A_{circ}$-asymmetry,
 for various choices of the (${\bar d_{B}} c_{W}^{2}+\bar
d s_{W}^{2}$) combination using: (a) $P_e=\bar P_e=-P_\gamma=- \bar
P_\gamma=\pm 1 $, $x_0=4.82$ and $\sqrt{s_{ee}}=\mh/0.75$; (b) 
$P_e=\bar P_e= P_\gamma= \bar P_\gamma=\pm 1 $, $x_0=4$ and
$\sqrt{s_{ee}}=0.5TeV$.\\
}
\label{comparison}
\end{figure}

\clearpage
\newpage

\vspace{-4cm}
\begin{center}
\[
\epsfxsize=7cm
\epsffile[57 135 576 715]{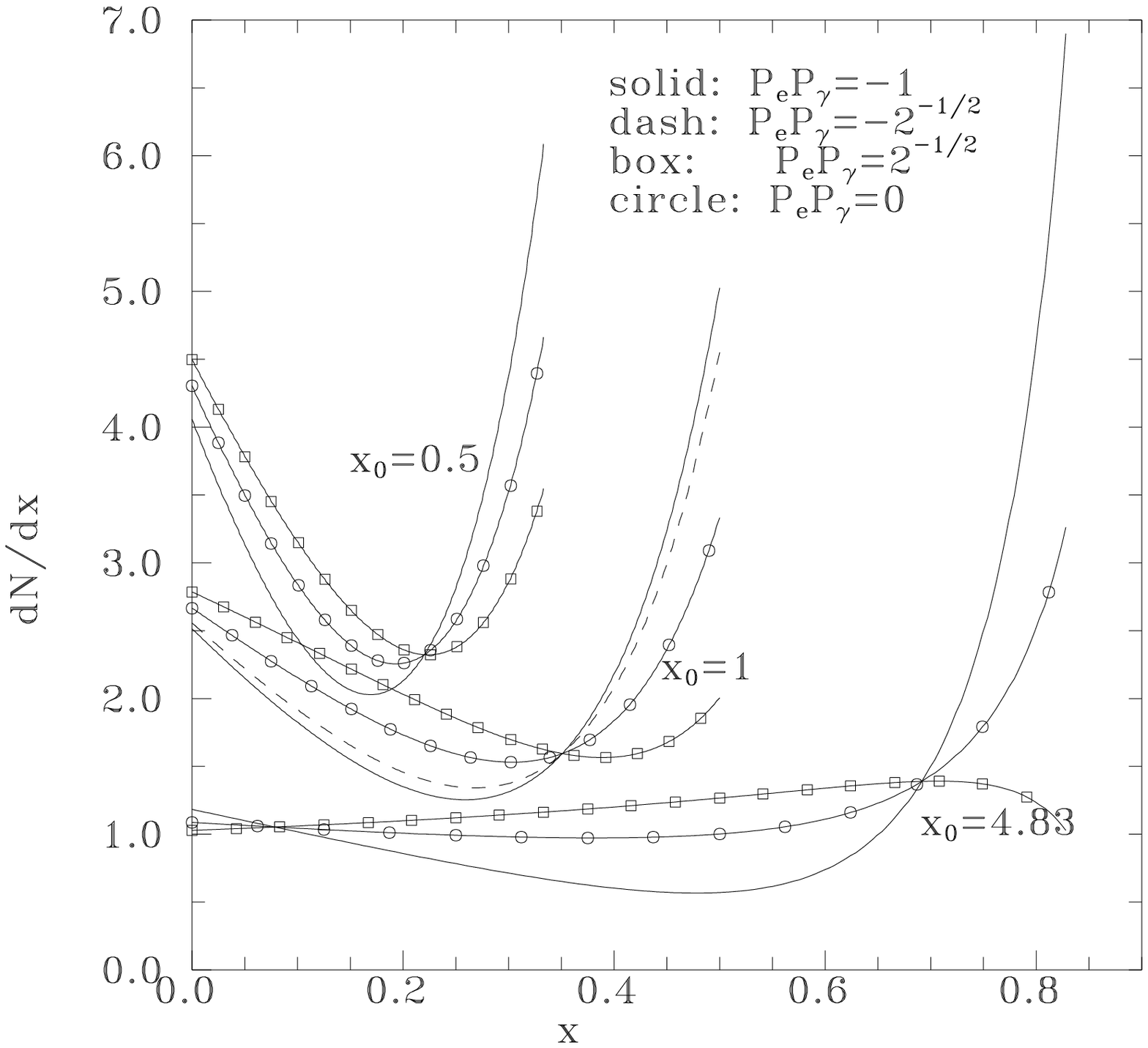}
\ \ \ \ \ \ \ \ 
\epsfxsize=7cm
\epsffile[57 135 576 715]{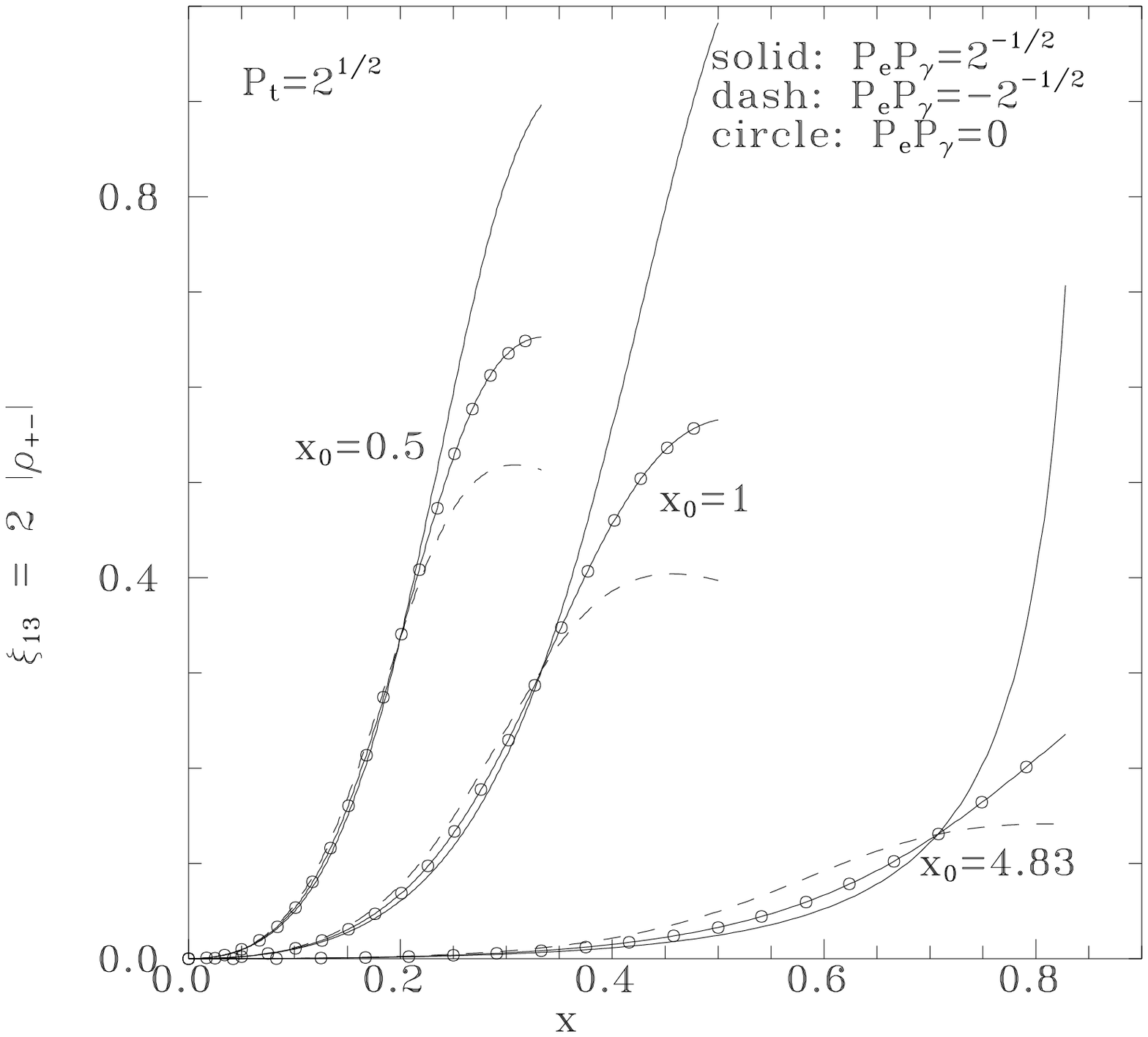}\]
\end{center}
\vspace{0.5cm}
\begin{center}
 (a) \hspace{8cm}  (b)
\end{center}
\vspace{-4cm}
\begin{center}
\[
\epsfxsize=7cm
\epsffile[57 135 576 715]{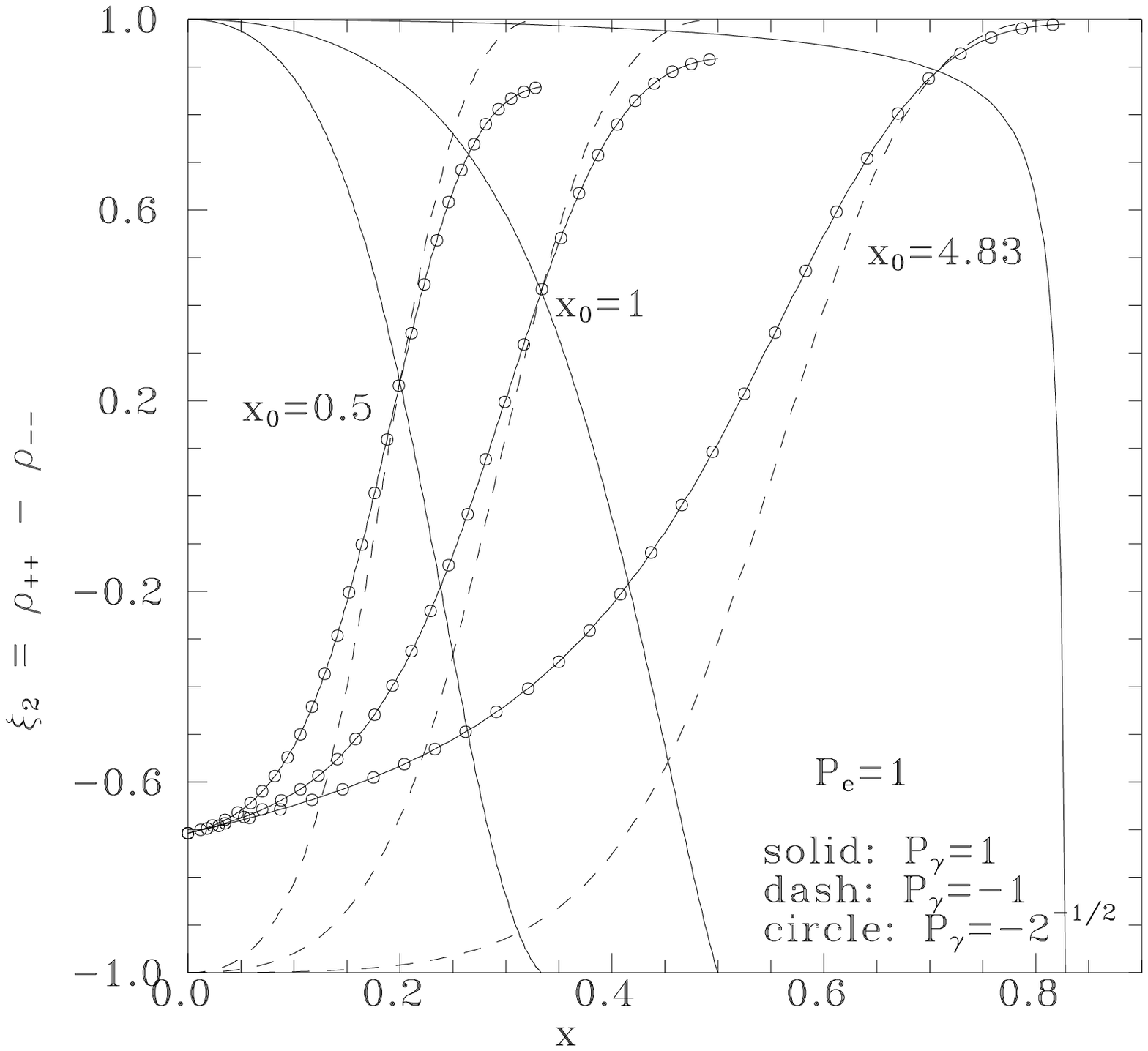}
\ \ \ \ (c) \]
\end{center}
\vspace{2cm}
\begin{center}
Figure 3: The spectrum of the backscatterd photon; (a) overall
flux, (b) average linear polarization of the backscattered
photon along the direction it is 
maximized, and (c) average circular polarization.
\end{center}

\end{document}